\documentclass[11pt]{article}
\usepackage{amsmath, amsfonts, amssymb}
\newcommand{\sepAuthor}{0.6in}
\newcommand{\sepAbstract}{0.6in}
\newcommand{\skipKeywords}{60pt}
\newcommand{\comments}[1]{}
\usepackage{amsthm}
\theoremstyle{plain}
\newtheorem{theorem}{Theorem}[section]
\newtheorem{lemma}[theorem]{Lemma}
\newtheorem{corollary}[theorem]{Corollary}

\theoremstyle{definition}
\newtheorem{definition}[theorem]{Definition}

\theoremstyle{remark}
\newtheorem{remark}[theorem]{Remark}
\newtheorem{reduction}[theorem]{Reduction}
\newcommand{\beginproof}{\medskip\noindent{\bf Proof.~}}

\newcommand{\finishproof}{\hspace{0.2ex}\rule{1ex}{1ex}}
\comments{
\newenvironment{proof}{\beginproof}{\unskip\nolinebreak\finishproof\par\medskip }
}

\def\squareforqed{\hbox{\rlap{$\sqcap$}$\sqcup$}}
\def\qed{\ifmmode\squareforqed\else{\unskip\nobreak\hfil
\penalty50\hskip1em\null\nobreak\hfil\squareforqed
\parfillskip=0pt\finalhyphendemerits=0\endgraf}\fi}

\newenvironment{proofof}[1]{\begin{trivlist}%
\item[]{\flushleft\em Proof of #1. }}
{\end{trivlist}}
\long\def\mytitlepage#1#2#3#4{
        \thispagestyle{empty}
        \begin{center}
        {\Large\bf #1}

        \vspace{\sepAuthor}
        #2\\
        \medskip

        \vspace{\sepAbstract}
        {\Large Abstract}
        \end{center}

        \noindent{#3}
        \vskip\skipKeywords

        \noindent{#4}
        \clearpage
        }
\newcommand{\bstr}[1]{\mbox{$\{0, 1\}^{#1}$}}

\newcommand{\ket}[1]{\mbox{$\left|{#1}\right\rangle$}}

\newcommand{\lb}[1]{\mbox{$\Omega\left({#1}\right)$}}
\newcommand{\up}[1]{\ensuremath{O\left({#1}\right)}}

\newcommand{\op}[1]{{\mathsf{#1}}}

\newcommand{\card}{\op{card}}
\newcommand{\set}[1]{\ensuremath{\left\{ {#1} \right\}}}

\newcommand{\norm}[1]{\ensuremath{\left\|{#1}\right\|}}

\newcommand{\E}{\ensuremath{\mathrm{E}}}

\newcommand{\domain}{\ensuremath{\mathrm{dom}}}
\newcommand{\image}{\ensuremath{\mathrm{img}}}

\newcommand{\rand}[1]{\mbox{\boldmath $#1$}}

\newcommand{\err}{\ensuremath{\epsilon}}

\newcommand{\edcite}{Buhrman, D\"urr, Heiligman, H{\o}yer,
Magniez, Santha, and de Wolf~\cite{BuhrmanDHHMSW01}}
\newcommand{\Bbcmw}{Beals, Buhrman, Cleve, 
Mosca, and de Wolf~\cite{BealsBCMW98}}

\newcommand{\allF}{\ensuremath{\mathcal{F}}}
\newcommand{\allPF}{\ensuremath{{\mathcal{F}^*}}}
\newcommand{\pF}{\ensuremath{{f^*}}}
\newcommand{\SG}{\ensuremath{SG}}
\newcommand{\pmf}{\ensuremath{\Gamma}}
\newcommand{\random}[1]{\mbox{\boldmath $#1$}}
\newcommand{\DTwoOne}{\ensuremath{D_{2\to1}}}
\newcommand{\DHTwoOne}{\ensuremath{D^{1/2}_{2\to1}}}
\newcommand{\DROne}{\ensuremath{D_{r\to1}}}
\newcommand{\DHROne}{\ensuremath{D^{1/2}_{r\to1}}}
\newcommand{\symmetrize}[1]{\ensuremath{\bar{#1}}}
\newcommand{\Qmg}[2]{\ensuremath{\symmetrize{P}(f_{#1,\, #2})}}
\begin{document}
\mytitlepage{
Quantum lower bounds 
for the collision and the 
element distinctness problems
      \footnote{
This work was supported in part by the National Science
Foundation under Grant. No. CCR-9820855, EIA-0086038,
and CCR-0049092.
}}
{{\large Yaoyun Shi}\\
\vspace{3ex}
{\large
    Institute for Quantum Information\\
    California Institute of Technology\\
    Pasadena, CA 91125, USA\\ 
    E-mail: shiyy@cs.caltech.edu
}}
{Given a function $f$ as an oracle, 
the collision problem 
is to find two distinct inputs $i$ and $j$ such
that $f(i)=f(j)$, under the promise
that such inputs exist.
Since the security of many fundamental 
cryptographic primitives depends on
the hardness of finding collisions,
quantum lower bounds for 
the collision problem
would provide evidence for the existence
of cryptographic primitives that are immune to 
quantum cryptanalysis.

In this paper, we prove that 
any quantum algorithm for 
finding a collision in an $r$-to-one
function must evaluate the function $\lb{(n/r)^{1/3}}$ times, 
where $n$ is the size of the domain and $r|n$.
This improves
the previous best lower bound of $\lb{(n/r)^{1/5}}$ 
evaluations due to Aaronson~[quant-ph/0111102] ,
and is tight up to a constant factor.

Our result also implies a quantum lower bound of $\lb{n^{2/3}}$
queries to the inputs for the element distinctness problem, which
is to determine whether or not the given $n$ real 
numbers are distinct. The previous
best lower bound is $\lb{\sqrt{n}}$ queries
in the black-box model;
and $\lb{\sqrt{n}\log{n}}$ comparisons in the
comparisons-only model, due
to H{\o}yer, Neerbek, and Shi~[ICALP'01, quant-ph/0102078].
}

\section{Introduction and summary of results}
The exponential speed-up of Shor's quantum 
algorithm for integer factorization~\cite{Shor94}
over the best known classical algorithm
has inspired scientists of many fields to
explore the power of quantum computing. 
On the other hand, understanding the limitations
of quantum computing is also of great importance.
Identifying problems that are hard for quantum
computers can not only deepen our knowledge
on the power of quantum computing, but is also
necessary for developing a new cryptography
immune to quantum cryptanalysis.

Given a function $f$ as an oracle, 
the collision problem 
is to find two distinct inputs $i$ and $j$ such
that $f(i)=f(j)$, under the promise
that such inputs exist.
This paper concerns the 
$r$-to-one collision problem, 
in which the oracle is promised to be $r$-to-one,
for some integer $r$ fixed in advance.
The case $r=2$ is important because
random two-to-one functions are considered 
good models of \emph{collision intractable functions},
which is a fundamental cryptographic primitive.
An exponential (in $\log n$)
quantum lower bound 
would be evidence for the existence
of collision intractable functions
for quantum computers.

Other motivations of our study arise from the
close connection of our problem
to other widely-studied problems. An example
is the \emph{hidden subgroup problem}, 
in which the input is some $r$-to-one function
with additional promises.
The Abelian case of the hidden subgroup problem
can be solved efficiently
by a natural generalization of the well-known quantum
algorithms of Simon~\cite{Simon97} and Shor~\cite{Shor97},
while the non-Abelian case is 
one of the major challenges in the design
of fast quantum algorithms 
(refer to 
Grigni, Schulman, Vazirani, and Vazirani~\cite{GrigniSVV01}
for a rencent development).
A quantum lower bound for Collision would
illuminate our understanding of the problem
structures that allow or disallow a quantum speed-up.

It is not hard to see that $\Theta(\sqrt{n/r})$
evaluations are sufficient and necessary
for classical algorithms to solve the $r$-to-one Collision.
Interestingly, quantum computers can do much
better:
using Grover's quantum search algorithm~\cite{Grover96}
in a novel way,
the quantum algorithm found by Brassard, H{\o}yer, 
and Tapp~\cite{BrassardHT97}
makes only $\up{(n/r)^{1/3}}$ evaluations.
Despite much research effort,
no lower bound better than constant 
had been found until very recently, when
Aaronson proved the ground-breaking 
$\lb{(n/r)^{1/5}}$ lower bound~\cite{Aaronson01}.
In this paper, we improve the lower bound to
the tight bound.

\begin{theorem}[Lower bound for Collision] 
\label{thm:mainrs}
Let $n>0$ and $r\ge2$ be integers
with $r|n$, and let a function of
domain size $n$ be
given as an oracle with
the promise that it is either
one-to-one or $r$-to-one.
Then any error-bounded quantum algorithm 
for distinguishing these two cases 
must evaluate the function
$\lb{(n/r)^{1/3}}$ times.
Thus, finding a collision in an
$r$-to-one function of domain size $n$
requires $\lb{(n/r)^{1/3}}$ evaluations.
\end{theorem}

Denote the set $\set{1, 2, \cdots, n}$ by $[n]$.
It remains an open problem whether or not
our lower bound
still holds if the range of oracle is restricted to
$[n]$. This is because Theorem~\ref{thm:mainrs}
is proved by considering oracles with range
$[3n/2]$.  Nevertheless, for the small range case, we are able
to improve Aaronson's $\lb{(n/r)^{1/5}}$ 
lower bound~\cite{Aaronson01} to $\lb{(n/r)^{1/4}}$.

\begin{theorem}[Lower bound for Collision with small range]
\label{thm:special} Let $n>0$ and $r\ge2$ be
integers with $r|n$, and a function
from $[n]$ to $[n]$ is given as an oracle
with the promise that it is either one-to-one
or $r$-to-one. Then any quantum algorithm
for distinguishing these two cases must evaluate
the function $\lb{(n/r)^{1/4}}$ times. 
Thus, finding a collision in an $r$-to-one
function from $[n]$ to $[n]$ must evaluate
the function $\lb{(n/r)^{1/4}}$ times.
\end{theorem}

Given $n$ real numbers,
are they all distinct? This is the 
classical problem of Element Distinctness, 
studied by many authors in the classical setting.
A simple algorithm would be to
sort the numbers using $\Theta(n\log{n})$ comparisons,
and then check the equality of neighboring numbers.
This is essentially optimal classically,
as suggested by the
many $\lb{n\log{n}}$ lower bounds
in various classical models.
In contrast, with another creative use
of Grover's algorithm~\cite{Grover96},
the quantum algorithm found by \edcite 
makes
only $O(n^{3/4}\log{n})$ comparisons.
Collision and Element Distinctness
are closely related, as we can see
from the following well-known reduction:
\begin{reduction}[From Two-to-one Collision to
Element Distinctness]
Run the algorithm for Element Distinctness
on the restriction of the oracle function on
a random set of $\Theta(\sqrt{n})$ inputs.
If the oracle is two-to-one, 
a collision will
be found with high probability, by the Birthday
Paradox.
\end{reduction}
Therefore, Theorem~\ref{thm:mainrs} implies,
\begin{corollary}[Lower bound for Element Distinctness]
\label{co:ele}
Any quantum algorithm that accesses the inputs through
an oracle and solves the element distinctness
problem of $n$ real numbers 
must make $\lb{n^{2/3}}$ oracle queries.
If only comparisons are allowed, the same number of
comparisons are required.
\end{corollary}

The previous best known quantum lower bound 
is $\lb{\sqrt{n}}$ queries to the inputs,
which can be obtained 
by a simple reduction from 
the search problem; and $\lb{\sqrt{n}\log{n}}$
comparisons in the comparisons-only model, 
due to H{\o}yer, Neerbek, and Shi~\cite{HoyerNS01}.
The gap between our lower bound and
the $\up{n^{3/4}\log{n}}$ upper bound of
Buhrman et al.~\cite{BuhrmanDHHMSW01}
remains to be closed.
The strongest classical lower bound is
the $\lb{n\log{n}}$ lower bound on the depth
of randomized algebraic decision trees, due
to Grigoriev, Karpinski, Meyer auf der Heide,
and Smolensky~\cite{GrigorievKHS96}.
For classical lower bounds in weaker models
refer to the papers by Ben-Or~\cite{BenOr83},
Steele and Yao~\cite{SteeleY82}, and, Dobkin and
Lipton~\cite{DobkinL78}. 
\comments{
element distinctness
is also studied in the classical space-time
trade-off setting. We will not discuss this
aspect, for the recent development of which
refer to the paper of Beame, Saks, Sun, and
Vee~\cite{BeameSSV00}.
}
\comments{
Other motivations for our study arise
from the close connections of the above problems
with some other well known problems. An example
is Graph Isomorphism, which is to
decide whether the given two graphs are isomorphic.
In the past few years, there has been much 
effort to find a polynomial time quantum algorithm
for Graph Isomorphism.
An exponential lower bound for the two-to-one
Collision would be a strong indication that
it can not be efficiently solved by quantum 
algorithms that only do simple search for collisions. 
For the recent development on
the quantum complexity of Graph Isomorphism and its
generalization, refer to the paper
of Grigni, Schulman, Vazirani, and Vazirani~\cite{GrigniSVV01}.
}

\begin{remark}
The worse-case and average-case
complexities of the collision problems
considered here are the same
because of their symmetry.
The reader may find it helpful
to regard the problems as bipartite
graph properties, and the inputs as bipartite
graphs. 
\comments{
Informally two functions are said to be 
\emph{isomorphic} if they are identical
after a permutation on the domain and a permutation
on the range.}
\end{remark}

\section{Proof outline and previous works}
\subsection{Proof outline}
From now on, we shall refer to 
distinguishing an $r$-to-one
function from a one-to-one function 
as the $r$-to-one problem, and denote it
by $\DROne$, or $\DROne(n, N)$
when the domain and range sizes
are $n$ and $N$, respectively.
For simplicity, we shall deal with 
$r=2$ in this section.

Our proof for Theorem~\ref{thm:mainrs}
takes two steps: first we reduce to 
$\DTwoOne$ a new problem \emph{Half-two-to-one},
which is then shown
to have an $\lb{n^{1/3}}$ lower bound.
Denote the set $\set{\frac{n}{2}+1, \frac{n}{2}+2, \cdots, n}$ by 
$[\frac{n}{2}+]$ ($n$ is even).

\begin{definition} Let $n>0$ be an integer and $4|n$.
In the {\bf half-two-to-one problem},
or $\DHTwoOne(n, n)$ for short, 
a function from $[n]$ to $[n]$
is given as an oracle with
the promise that half of the inputs
are two-to-one mapped to $[\frac{n}{2}+]$,
and the other half are mapped to
$[\frac{n}{2}]$, either one-to-one or
two-to-one. The problem is to distinguish
these two cases.
\end{definition}

\begin{lemma}\label{lm:reduction}
$\DHTwoOne(n, n)$ can be reduced to
$\DTwoOne(n, 3n/2)$ with a constant factor slow-down.
\end{lemma}

\begin{theorem}\label{thm:mainlb}
Any quantum algorithm for $\DHTwoOne(n, n)$
requires $\Omega(n^{1/3})$ evaluations.
\end{theorem}

The reduction is done by exploring
the symmetry of the problems, and
by using the following important fact:
on the $n/2$ inputs mapped
to $[\frac{n}{2}+]$, $f$ can be modified
to be one-to-one mapped to $[3n/2]\backslash[n/2]$,
without much slow-down.

We prove Theorem~\ref{thm:mainlb} by
using the polynomial method of \Bbcmw, and
Aaronson~\cite{Aaronson01} with new ideas.
More specifically, let us fix a
$T$-queries algorithm $\mathcal{A}$
for $\DHTwoOne(n, n)$.
First we \emph{symmetrize}
$\mathcal{A}$ to obtain $\symmetrize{\mathcal{A}}$
so that running $\symmetrize{\mathcal{A}}$
on any input $f$ is equivalent to running $\mathcal{A}$
on a random input $\rand{f}$ ``isomorphic'' to $f$.
Then we run $\symmetrize{\mathcal{A}}$ on the oracle
$f_{m, g}$, the function $g$-to-one mapped to $[n/2]$
on the first $m$ inputs and two-to-one
mapped to $[\frac{n}{2}+]$ on the remaining.

Following an important observation of 
Beals et al.~\cite{BealsBCMW98}
that relates the number of quantum queries
to polynomial degrees, and
from the nice symmetry of
$\symmetrize{\mathcal{A}}$, 
the acceptance probability $\symmetrize{P}(f_{m, g})$
turns out to be a polynomial in $m$ and $g$ 
with degree $\le 2T$. In addition, 
for all $m$ and $g$ such that $f_{m, g}$ is well-defined,
$\symmetrize{P}(f_{m, g})\in[0, 1]$; and, there is a gap between
$\symmetrize{P}(f_{\frac{n}{2}, 1})$ and 
$\symmetrize{P}(f_{\frac{n}{2}, 2})$.
These two nice properties enable one
to apply a theorem by Paturi~\cite{Paturi92}
to prove the desired lower bound
for $\deg(\Qmg{m}{g})$. We point out that
essentially
Paturi's theorem follows from
both Markov Inequality 
and Bernstein Inequality, 
two fundamental theorems in approximation theory that
give good lower bounds for polynomial degrees.

For proving Theorem~\ref{thm:special},
we need the following additional idea.
Given an algorithm for
$\DTwoOne(n, n)$, we modify the algorithm
so that it can be run
on inputs that are
only \emph{partially} defined:
Whenever the algorithm queries
an undefined input,
we force the algorithm to \emph{abort}
on the corresponding base vector.
The rest of the proof is similar to that
for Theorem~\ref{thm:mainrs}.

\begin{remark}Running the symmetrized 
algorithm on a fixed input is equivalent to
running the algorithm on some random input,
as treated by Aaronson~\cite{Aaronson01}.
However, we feel that our treatment 
explores the symmetry of the problem more
explicitly and thus makes it less mysterious 
that the acceptance probability turns out to be
a polynomial.
\end{remark}

\subsection{Relation with previous works}
Aaronson~\cite{Aaronson01}
introduces the following original 
lower bound idea, which we shall refer to as the 
\emph{derived polynomial method}:
run the given $T$-queries algorithm on 
$\random{f}_y$, a probability distribution
determined by a parameter $y$.
A new polynomial on $y$
of $O(T)$ degree is derived from the average
acceptance probability, and
is then shown to have high degree by other
methods.
He is also the first to consider running
the given algorithm on almost $g$-to-one
functions for arbitrary $g$.
We follow this approach in proving 
Theorem~\ref{thm:mainlb} and 
improve his proof in the following
ways:
(1) The derived polynomial
method seems to be more effective on
Half-two-to-one than on 
Two-to-one itself. 
This is because the structure of Half-two-to-one 
yields a polynomial that has a gap around
$m=n/2$, while the range of $m$ is $[0, n]$.
This feature, lacking in \cite{Aaronson01},
is very important because it allows
one to apply Bernstein Inequality, which
in general gives a better degree lower bound than
Markov Inequality if the function value has
a sudden change close to the center of the domain.
(2) The corresponding input distributions
in our proofs are more natural and effective.
As consequences, not only the ranges
of the parameters are larger, but also
the acceptance probabilities are exactly polynomials,
instead of being close to a polynomial 
as in~\cite{Aaronson01}.
Thus better lower bounds can be obtained
with simpler algebra.

It seems to us that our partial input idea was not
used before. Another novel way of manipulating inputs for 
proving quantum lower bounds is used by
Ambainis~\cite{Ambainis99},
where an adaptive adversary changes the input according
to the performance of the algorithm.

Our problem can be formulated
in the \emph{black-box computation model},
a model widely studied in recent years
due to both its simplicity 
and its power in modeling many natural problems. 
\comments{
Besides the polynomial method,
two other general lower bound techniques for
this model were developed:
the hybrid method, introduced
by Bennett, Bernstein, Brassard, and Vazirani~\cite{BennettBBV97};
and the adversary method, introduced by
Ambainis~\cite{Ambainis00}.
The recent paper by Ambainis~\cite{Ambainis01} 
is an excellent survey on black-box quantum
computation with an emphasis on lower bounds.
}
For other techniques for proving quantum lower bounds
in this model, and quantum black-box computation in general,
refer to the excellent survey of
Ambainis~\cite{Ambainis01}.

We remark that previous approaches for proving
degree lower bounds for (partial) Boolean functions
can be interpreted in the light of 
the derived polynomial method.
For example, the \emph{symmetrization method}, introduced 
by Minsky and Papert~\cite{MinskyP69}
and used by Paturi~\cite{Paturi92}
and Nisan and Szegedy~\cite{NisanS92},
symmetrizes a Boolean function
uniformly over all permutations of the Boolean variables.
Another example, the \emph{linear approximation}
technique used by Shi~\cite{Shi01b},
averages a Boolean function
by tossing independent coins for each Boolean
variable, and the mean value of each coin
is a linear function of a single parameter.

The rest of this paper is organized as follows.
In Section~\ref{sec:pre}, we define the black-box model,
introduce some notations, and state
theorems from approximation theory which our 
proofs finally rely on
the theorem of Paturi~\cite{Paturi92}.
We then prove our lower bound for the general case
of Collision in Section~\ref{sec:mainrs}, which is followed by
the proof for the special case of small range.
Finally, we discuss some open problems.

\section{Preparations for the proofs}
\label{sec:pre}
Let $n\ge 0$ and $N\ge 0$ be integers and
$\allF:=\allF(n, N)$ be the set of all functions from $[n]$ to
$[N]$. Let $f\in\mathcal{F}$ be given as an oracle.
Following Beals et
al.~\cite{BealsBCMW98}, we give the following 
definition of the black-box model, customized to our setting.

A quantum black-box algorithm works in a Hilbert space
of dimension $n^2 L$, for some $L:=L(n)<+\infty$.
An orthonormal basis is chosen and denoted by
\[ \set{\ket{i}\ket{j}\ket{l}: i, j\in[n], l\in[L]}.\]
For $j\in[N]$ and $j'\in[N]$, 
define $j + j' \mod N:= i+j - \lfloor (i+j+1)/N \rfloor \cdot N$.
An oracle gate is the following unitary operator determined by
$f$:
\[ \op{O}_f \ket{i, j, l} := \ket{i,\ f(i) + j \mod N,\ l},\quad\quad \forall
i\in[n], j\in[N], l\in[L].\]
A quantum black-box algorithm that makes $T$ queries consists
of $T+1$ unitary operators, $\op{U}_0, \op{U}_1, \cdots,
\op{U}_T$, and a projection operator $\op{P}$, on the Hilbert space.
It starts with a constant vector 
denoted by $\ket{0}$,
then applies the following sequence of operators:
\[ \op{U}_0 \rightarrow \op{O}_f \rightarrow \op{U}_1
\rightarrow\cdots\rightarrow\op{U}_{T-1}\rightarrow
\op{O}_f\rightarrow\op{U}_T\rightarrow\op{P}.\]
The acceptance probability is
\[ P(f) := \left\| \op{P}\op{U}_T\op{O}_f\op{U}_{T-1}\cdots\op{O}_f
\op{U}_0\ket{0}\right\|^2.\]

We say that the algorithm computes 
a function $\phi:\allF\supseteq\allF'\rightarrow\bstr{}$,
where $\allF'\subseteq \allF$,
with error probability bounded by
$\err$ if for every $f\in \allF'$, $|P(f) - \phi(f)| \le \err$.
The quantum complexity of $\phi$ is the minimal integer $T$ such
that there exists a quantum algorithm that computes $\phi$ with
$T$ queries and errs with a probability bounded by $1/3$.

As before, for all $i\in[n]$ and $j\in[N]$, the predicate
$\delta_{i, j}(f):=1$ if and only if
$f(i) = j$.
Observe that for all $i$, $j$, $l$, and $f$,
\[ \op{O}_f \ket{i, j, l} = \sum_{j'=1}^{N}\, \delta_{i, j'}(f)
\, \ket{i,\ j + j' \mod N,\ l}.\]
Since all $\op{U}_t$ and $\op{P}$ are linear transformations,
we have the following important observation
by Beals et al.~\cite{BealsBCMW98}, in the form stated
in Aaronson~\cite{Aaronson01}:

\begin{lemma}
\label{lm:poly}
The acceptance probability $P(f)$ can be expressed
as a polynomial over the predicates
$\delta_{i,j}$, $i\in[n], j\in[N]$, and
$\deg(P)\le 2T$.
\end{lemma}

Let $\allPF:=\allPF(n, N)$ denote
the set of all \emph{partial} functions from
$[n]$ to $[N]$.
Denote the domain and image of a function
$\pF$ by $\domain(\pF)$ and $\image(\pF)$, respectively.
Any $\pF\in\allPF$ can be
conveniently represented as a subset
of $[n]\times[N]$, i.e.,
$\pF = \set{(i, \pF(i)) : i\in\domain(\pF)}$.
For a finite set $K\subseteq\mathbb{Z}^+$,
let $\SG(K)$ denote the group of permutations on $K$.
Any permutation in $\SG(K)$ is 
understood as the identity mapping on 
any $k'\notin K$.
For any integer $k>0$, $\SG(k)$ is a shorthand
for $\SG([k])$. 
For each $\sigma\in \SG(n)$ and $\tau\in\SG(N)$,
define $\pmf_\tau^\sigma: \allPF\rightarrow \allPF$
as
\[\pmf^\sigma_\tau(\pF) := 
\set{(\sigma(i), \tau(j)): (i, j)\in \pF}, \qquad
\forall \pF\in \allPF\,.\]

For all $s\in\allPF$,
the predicate $I_{s}:\allPF\rightarrow\bstr{}$
is defined as follows:
\[I_{s}(\pF):= 1 \iff
s\subseteq \pF,
\qquad\forall \pF\in\allPF.\]

Fix a quantum black-box algorithm that
queries $T$ times. By Lemma~\ref{lm:poly},
the acceptance probability can be written as
\begin{equation}
\label{eqn:poly}
 P(f) = \sum_{s \in \allPF,
  \card(s)\le 2T} \beta_{s}
\  I_s(f),
\quad\quad\quad \forall s,\ \beta_s\in \mathbb{R}.
\end{equation}

Now proving a quantum lower bound is reduced
to proving a lower bound on $\deg(P)$, for which
we will resort to the following two fundamental
theorems from approximation theory.
For any function $q:\mathbb{R}\rightarrow\mathbb{R}$,
and any set $D\subseteq\mathbb{R}$,
let $\norm{q}_D$ denote $\sup\set{|q(\alpha)|:\alpha\in D}$.

\begin{theorem}[Markov Inequality]
\label{thm:markov} For any polynomial 
$q(\alpha)\in\mathbb{R}[\alpha]$ with degree $d$
and $\norm{q}_{[-1, 1]} = 1$,
\[ \norm{q'}_{[-1, 1]} \le d^2.\]
\end{theorem}

\begin{theorem}[Bernstein Inequality]
\label{thm:bernstein} For any polynomial
$q(\alpha)\in\mathbb{R}[\alpha]$ with degree $d$
and $\norm{q}_{[-1, 1]} = 1$,
\[ |q'(\alpha)| \le \frac{d}{\sqrt{1-\alpha^2}},
\quad\quad\forall \alpha\in(-1, 1).\]
\end{theorem}

The proofs for the above theorems 
can be found in Chapter~4 of the book by 
Devore and Lorentz~\cite{DevoreL93}.
We will actually use the following result
that follows from the above theorems.
It is proven (with slight modification)
by Paturi~\cite{Paturi92}
in giving tight bounds for the lowest
degree polynomial approximation to symmetric
Boolean functions.

\begin{theorem}[Paturi~\cite{Paturi92}]
\label{thm:paturi} 
 Let $q(\alpha)\in\mathbb{R}[\alpha]$
be a polynomial of degree $d$,
 $a$ and $b$ be integers with $a<b$, 
and $\xi\in[a, b]$ be a real number.
If
(1) $|q(i)|\le 1$ for all integers $i\in[a, b]$; 
and,
(2) $\left|q(\lfloor\xi\rfloor) - q(\xi)\right|\ge c$
for some constant $c>0$.
Then,
\[ d = \lb{\sqrt{(\xi-a+1)(b-\xi+1)}}.\]
In particular,
\[ d = \lb{\sqrt{b-a}}.\]
\end{theorem}

As a convention, all random variables are
uniform over their domain.

\section{Lower bound for the general collision problem}
\label{sec:mainrs}
\subsection{The reduction}
\begin{proofof}{Lemma~\ref{lm:reduction}}
Let $\mathcal{A}$ be a quantum algorithm
for $\DTwoOne(n, 3n/2)$.
We shall derive an algorithm
$\mathcal{B}$ for $\DHTwoOne(n, n)$.

We call a function $f$ half-two-to-one,
if it is one-to-one on a half of its input,
two-to-one on the other half,
and, the two images  are disjoint.
Let $p_1\ge2/3$, $p_0\le1/3$, and $p_{1/2}$ be the acceptance 
probabilities
of $\mathcal{A}$ with the input being a random
two-to-one, one-to-one, and half-two-to-one
function from $[n]$ to $[3n/2]$, respectively.
Let $f$ be the oracle function for the 
$\DHTwoOne(n, n)$ problem. Then $f$
is either half-two-to-one or two-to-one,
with some additional constrains on the range.

If $p_{1/2}<1/2$, $\mathcal{B}$ will be the following:
Choose random variables $\random{\sigma}\in\SG[n]$
and $\random{\tau}\in\SG[3n/2]$, then
run $\mathcal{A}$ on 
$\random{f}:=\pmf_{\random{\tau}}^{\random{\sigma}}(f)$.
If $f$ is two-to-one, the algorithm
will accept with probability $p_1\ge 2/3$;
otherwise it will
accept with probability $p_{1/2}<1/2$.

Assume $p_{1/2} \ge 1/2$. 
Define $\bar{f}:[n]\rightarrow[3n/2]$ as:
\begin{equation}
\bar{f} (i) := \begin{cases}
            i + n/2 & \text{if $f(i)>n/2$;}\\
            f(i) & \text{otherwise.}
                \end{cases}
\end{equation}
Notice that the oracle $\op{O}_{\bar{f}}$ can be simulated
by two applications of $\op{O}_f$ together with
some local unitary operators.
Now $\mathcal{B}$ will be:
Choose random variables 
$\random{\sigma}\in\SG(n)$, and
$\random{\tau}\in\SG(3n/2)$,
then run $\mathcal{A}$ on 
$\random{f}:=\pmf_{\random{\tau}}^{\random{\sigma}}(\bar{f})$.

Note that
for each $i$ with $f(i)\in[\frac{n}{2}+]$,
$\bar{f}(i)$ is a distinct number in 
$[3n/2]\backslash[n/2]$.
Therefore, if $f$ is half-two-to-one,
$\bar{f}$ is one-to-one, 
in which case $\random{f}$ is a random one-to-one function;
thus $\mathcal{B}$ will accept
with probability $p_0\le 1/3$.
On the other hand, if $f$ is two-to-one, 
$\bar{f}$ is half-two-to-one,
in which case $\random{f}$ is a random half-two-to-one function;
thus $\mathcal{B}$ will accept
with probability $p_{1/2}\ge 1/2$.
\qed
\end{proofof}

\subsection{Lower bound for the half-two-to-one problem}
Fix a quantum algorithm for $\DHTwoOne(n, n)$,
and let $P(f)$ be its acceptance probability.
To prove an $\lb{n^{1/3}}$ lower bound
for $\DHTwoOne(n, n)$, we need only to prove
the lower bound for $\deg(P)$, by Lemma~\ref{lm:poly}.
Define the symmetrization of $P$ as
\begin{equation}
\label{eqn:sym}
 \symmetrize{P}(f) :=
\E_{\random{\sigma}\in\SG(n), \random{\tau}\in\SG([n/2]), 
          \random{\tau'}\in\SG([\frac{n}{2}+])}
\left[\ P\left(
   \pmf_{\random{\tau} \circ \random{\tau'}}^{\random{\sigma}}(f)
          \right)
\ \right].
\end{equation}

\begin{definition}
\label{def:valid}
We call a pair of integers $(m, g)$
{\bf valid}, if
$0\le m \le n$, $1\le g\le n$,
$2|m$, $g|m$, and if $g=1$, $m\le n/2$.
\end{definition}

Given a valid $(m, g)$,
define $f_{m, g}:[n]\rightarrow[n]$ as follows:
\begin{equation}
f_{m, g}(i) = \begin{cases}
     \lceil i/g\rceil & i\in[m],\\
     \lceil(i-m)/2\rceil + n/2 &\text{otherwise.}
     \end{cases}
\end{equation}
%Let $\symmetrize{P}(f_{m, g}) := \symmetrize{P}(f_{m, g})$.

\begin{lemma}
\label{lm:main} The function $\symmetrize{P}(f_{m, g})$
is a polynomial in $m$ and $g$ of degree $\le 2T$. 
\end{lemma}

\begin{proof} 
By Lemma~\ref{lm:poly}, it suffices to show that for each monomial
$I_s$, $\card(s)\le2T$,
the symmetrization $\symmetrize{I}_s$ is such a polynomial, where
\[\symmetrize{I}_s(f_{m, g}):=\E_{\random{\sigma}, \random{\tau}, 
          \random{\tau'}}
\left[\ I_s\left(
   \pmf_{\random{\tau} \circ \random{\tau'}}^{\random{\sigma}}(f_{m, g})
          \right)
\ \right].\]

Let $w: = \card(\image(s)\cap[n/2])$.
Fix a sequence of elements in $\image(s)\cap[n/2]$,
and let $u_1, u_2, \cdots, u_w$ be the 
corresponding sequence of sizes of preimages for
the elements.
Put $u:=\sum_{j=1}^w u_j$.
Replacing $[n]$ by $[\frac{n}{2}+]$, we define
$w'$, $u'_j$, $1\le j \le w'$, and $u'$,
similarly.
For integers $a, b$, $P_a^b := a(a-1)\cdots(a-b+1)$.
Put
\[ \lambda:= \frac{(n/2-w)!(n/2-w')!(n-u-u')!}{n!(n/2)!(n/2)!}.\]
By simple calculations, 
\begin{eqnarray}
\symmetrize{I}_s(f_{m, g})
&=&
\lambda\ \cdot
\ P_{m/g}^w\cdot\Pi_{j=1}^{w} P_g^{u_i}\cdot
P_{\frac{n-m}{2}}^{w'}\cdot
     \Pi_{j=1}^{w'} P_2^{u'_i}\\
&=&\lambda\cdot
\Pi_{j=0}^{w-1}(m-g\cdot j) \cdot
\Pi_{j=1}^{w} P_{g-1}^{u_i-1}
\cdot\Pi_{j=0}^{w'-1} (n-m-2j)
\cdot \Pi_{j=1}^{w'} P_1^{u'_i-1},
\end{eqnarray}
which is a polynomial in $m$ and $g$ of 
degree
\[w + (u-w) + w' + (u' - w') = u+u' = \card(s) \le 2T.\]
\end{proof}

\begin{proofof}{Theorem~\ref{thm:mainlb}}
Since $\deg(\symmetrize{P}(f_{m, g})) \le 2T$ by the above lemma,
it suffices to prove $\deg(\Qmg{m}{g}) = \lb{n^{1/3}}$.

Since $\symmetrize{P}(f_{m, g})$ is defined to be the
acceptance probability for the oracle $f_{m, g}$,
\begin{equation}
\label{eqn:valid}
0\le \symmetrize{P}(f_{m, g}) \le 1,\quad
\textrm{for all valid $(m, g)$,}
\end{equation}
\begin{equation}
\label{eqn:jump}
0\le \Qmg{n/2}{1} \le 1/3,
\qquad\qquad
\textrm{and,}
\qquad\qquad
2/3\le \Qmg{n/2}{2} \le 1.
\hspace{\stretch{1}}
\end{equation}

Put $G:=\lfloor n^{2/3} \rfloor$, and
$Q_1(\alpha) := \Qmg{n/2}{\alpha}$.
Clearly, $\deg(Q_1)\le \deg(\Qmg{m}{g})$.
By Equations in~\ref{eqn:jump},
\[ \left|Q_1(1) - Q_1(2)\right| = 
\left|\Qmg{n/2}{1} - 
\Qmg{n/2}{2}\right| \ge 1/3.\]
If $|Q_1(k)|\le 2$ for all $k\in[G]$,
by Theorem~\ref{thm:paturi},
$\deg(Q_1) = \lb{\sqrt{G}}$,
which implies $\deg(\Qmg{m}{g})=\lb{n^{1/3}}$.
Otherwise, let $g_0\in[G]$ be such that
$|Q_1(g_0)| > 2$.

Put $G_0 := \lfloor\frac{n}{2g_0}\rfloor$,
and, 
\[Q_2(\alpha) := \Qmg{2 g_0 \alpha}{g_0},
\quad\quad \alpha\in[0, G_0].\]
Then $G_0 = \lb{n^{1/3}}$, and
 $\deg(Q_2)\le \deg(Q)$.
Since $g_0\ge 2$, 
$(2i g_0, g_0)$ is valid
for each $i\in[G^*_0]$, which implies
$0\le Q_2(i)\le 1$, by Eqn.~\ref{eqn:valid}.
Since
\[\left|Q_2\left(\frac{n}{4g_0}\right)\right| = 
\left|\Qmg{n/2}{g_0}\right| = 
\left|Q_1(g_0)\right| >2,
\]
and $0 \le Q_2\left(\left\lfloor
\frac{n}{4g_0}\right\rfloor\right)\le 1$, we have,
\[\left|Q_2\left(\left\lfloor\frac{n}{4g_0}\right\rfloor\right) - 
   Q_2\left(\frac{n}{4g_0}\right)\right| \ge 1.\]
Applying Theorem~\ref{thm:paturi}, we have 
\[\deg(Q_2)= \Omega\left(
\sqrt{\left(\frac{n}{4g_0}+1\right)\left(G_0 - \frac{n}{4g_0} +1\right)}
\right),\]
which implies
$\deg(\Qmg{m}{g}) = \lb{n^{1/3}}$.
\qed
\end{proofof}

\subsection{Generalizing to arbitrary $r\ge2$}
\begin{proofof}{Theorem~\ref{thm:mainrs}}
Combining Lemma~\ref{lm:reduction} and
Theorem~\ref{thm:mainlb} we obtain 
Theorem~\ref{thm:mainrs} for the case $r=2$.
To generalize to arbitrary $r\ge2$,
we need only to replace Half-two-to-one by
 \emph{Half-$r$-to-one}, denoted by 
$\DHROne(n, \frac{n}{2} + \frac{n}{r})$,
where the oracle 
is $r$-to-one mapped to $[n/2 + 1, n/2+2, \cdots, n/2 + n/r]$
on $n/2$ inputs 
and the other $n/2$ inputs are mapped to $[n/2]$
either $r$-to-one or one-to-one. 
In Definition~\ref{def:valid}, the condition
$2|m$ for $(m, g)$ being valid is replaced by $r|m$.

In analogy to Lemma~\ref{lm:reduction},
$\DHROne(n, \frac{n}{2} + \frac{n}{r})$ 
can be reduced to $\DROne(n, 3n/2)$.
To prove the $\lb{(n/r)^{1/3}}$ lower bound
for the former,
we need only to modify the proof
for the latter by choosing appropriate
parameters. That is, we set 
$G:=\left(\left\lfloor (n/r)^{2/3}\right\rfloor\right)\cdot r$.
We leave the remaining work for interested readers.
\qed
\end{proofof}

\section{Lower bound for Collision with small range}
\label{sec:special}
Let $n$ and $r$ be integers, and $r|n$. Fix
a $T$-queries 
quantum black-box algorithm for $\DROne(n, n)$.
Let $P(f)$ be its acceptance probability. 
Instead of making a reduction, we need the following lemma.

\begin{lemma}\label{lm:bounded}
 For any partial assignment $s$, 
\[ 0\le P(s) \le 1.\]
\end{lemma}

\begin{proof} Let $\op{U}_t$, $0\le t \le T$, and 
$\op{P}$ be the unitary operators and the final projection 
operator of the algorithm.  
Let $\op{P}_s$ be the operator that projects a state to the subspace
spanned by 
\[ \set{\ket{i, j, l}: i\in\domain(s), j\in[n], l\in[L]}.\]
Then it can be easily proved by induction that 
\[P(s) = \left\| \op{P}\op{P}_s \op{U}_T \op{O}_x \op{P}_s 
\cdots \op{P}_s \op{U}_1\op{O}_x \op{P}_s \op{U}_0 
\ket{0}\right\|^2.\]
The lemma follows.
\end{proof}

The symmetrization of $P$ is defined as
\[ \symmetrize{P}(f):=\E_{\rand{\sigma}, \rand{\tau}\in\SG(n)}
\left[ P(\pmf^{\rand{\sigma}}_{\rand{\tau}}(f)) \right].\]
Now we call a pair of integers $(m, g)$ 
\emph{valid} if $m\in[n^*]$, $g\in[n]$, and $g|m$.
Given a valid $(m, g)$, define the partial
function $f_{m, g}$ as follows:
\[ f_{m, g}:=\set{ (i, \lceil i/g \rceil) : i\in[m]}.\]

By Lemma~\ref{lm:bounded} and the definition of
$\symmetrize{P}(f_{m, g})$, 
\begin{equation}
\label{eqn:valid2}
0\le \symmetrize{P}(f_{m, g}) \le 1,\qquad
 \textrm{for all valid $(m, g)$.}
\end{equation}
By the correctness of the algorithm,
\begin{equation}
\label{eqn:jump3}
2/3 \le \Qmg{n}{1} \le 1, \qquad\qquad\textrm{and,}
\qquad\qquad 0\le \Qmg{n}{r} \le 1/3.
\end{equation}

\begin{lemma}
\label{lm:main2}
The function $\symmetrize{P}(f_{m, g})$ can be expressed as
a polynomial in $m$ and $g$ of degree $\le 2T$.
\end{lemma}

We omit the proof since it is in analogy to
the proof for Lemma~\ref{lm:main}.

\begin{proofof}{Theorem~\ref{thm:special}} By 
Lemma~\ref{lm:main2}, it suffices
to prove $\deg(\symmetrize{P}(f_{m, g})) = \Omega((n/r)^{1/4})$.
The proof is similar to that for 
Theorem~\ref{thm:mainlb}, and is much simpler.
We leave the details to the reader.
\qed
\end{proofof}

\section{Open problems}
Besides the two mentioned open problems,
Collision with small range 
and Element Distinctness,
we raise two more.

\begin{definition}
Two sets $f=\set{f(1), f(2), \cdots, f(n)}$
and $g=\set{g(1), g(2), \cdots, g(n)}$
are given as oracles with
the promise that either $f=g$ or $f\cap g=\varnothing$.
The {\bf set equality problem} is to distinguish these two cases.
\end{definition}

This is a special case of the
two-to-one problem, and it closely models the Graph Isomorphism
problem. We are not able to prove any
$\omega(1)$ lower bound, while we conjecture that
it is as hard as the general Collision.
A problem harder than the above is:

\begin{definition}
Given $n$ distinct numbers $x_1, x_2, \ldots, x_n$, 
the {\bf index erasure problem} is to
generate a vector close to
$\ket{\phi_x}=\frac{1}{\sqrt{n}}\sum_{i=1}^{n}\ket{x_i}$.
\end{definition}

This problem is equivalent to the following 
{\bf quantum-parallel search problem}:
Given an oracle described above, and the 
state $\ket{\phi_x}$,
generate a vector close to 
$\frac{1}{\sqrt{n}}\sum_{i=1}^n\ket{x_i}\ket{i}$.
One can show that $O(\sqrt{n})$ queries are sufficient 
for both problems by using Grover's quantum
search algorithm~\cite{Grover96}. 
We conjecture that this is tight,
though we are not able to prove any $\omega(1)$ lower bound.

\section{Acknowledgments}
I am most grateful to Ronald de Wolf and Scott Aaronson for 
stimulating discussions and valuable comments, and to 
Scott Aaronson for sending me his results
on Collision. Other thanks go to Lawrence Ip,
Ashwin Nayak, and Leonard Schulman for helpful discussions
and comments. 
I am also indebted to Andy Yao, Umesh Vazirani, and Ashwin Nayak
for introducing these problems to me and for
the numerous discussions at the early stage of this work.

\end{document}